\documentclass[a4paper,12pt]{article} 
\usepackage{graphicx} 
\usepackage{graphics} 

\usepackage[colorlinks=true, pdfstartview=FitV, linkcolor=blue,  citecolor=blue, urlcolor=blue]{hyperref}  %liens hypertexte

\def\l{\left}
\def\r{\right}
\def\be{\begin{equation}}
\def\ee{\end{equation}}
\def\beq{\begin{equation}}
\def\eeq{\end{equation}}
\def\d{\partial}

\begin{document}

\title{Quantum-like gravity waves and vortices in a classical fluid}
\author{Laurent Nottale\\{\small CNRS, LUTH, Paris Observatory and Paris-Diderot University,} \\{\small 5 place Janssen, 92190 Meudon, France}}
\maketitle

\begin{abstract}
We have recently proposed a new general concept of macroscopic quantum-type experiment. It amounts to transform a classical fluid into a quantum-type fluid by the application of a quantum-like potential, either directly in a stationary configuration, or through a retro-active loop to simulate the time evolution. In this framework, the amplitude of the quantum potential depends on a macroscopic generalization of the Planck constant, which can be changed during the experiment, therefore simulating a quantum to classical transition. The experiment is exemplified here by an application of this concept to gravity waves at the surface of an incompressible liquid in a basin of finite height, with particular emphasis on the quantized vortex. We construct a complex wave function with the height of the fluid in the basin as its square modulus and the velocity potential as its phase. This wave function is solution of a nonlinear Schr\"odinger equation typical of superfluids. The quantum potential is therefore defined here in terms of the square root of the fluid height. We suggest two methods for applying this quantum-like potential to the fluid: (i) by the action of a force on the surface (wind, blower, pressure, field, etc...); (ii) by a curvature of the basin ground. In this last case the ground profile yields the quantum potential itself, while usually only the quantum force is accessible, so that such an experiment is expected to provide one with a macroscopic model of a quantum-type vacuum energy. These results may also be relevant to the study of freak waves, which have already been described by nonlinear Schr\"odinger equations.
\end{abstract}

%1*****************
\section{Introduction}
%*****************
A new kind of macroscopic quantum-type experiment has recently been proposed \cite{LN06,NL06,LN07}. It consists of applying, through a real time retroactive loop, a generalized quantum-like potential on a classical system. Indeed, one can show that the Euler and continuity system of equations that describes a fluid in irrotational motion subjected to a generalized quantum potential is equivalent to a generalized Schr\"odinger equation \cite{LN07,LN93,LN97}. This is true also for a large class of rotational fluid motions, in which case one obtains a Schr\"odinger equation in a vectorial potential of the electromagnetic type \cite{LN07}. In this derivation, the quantum potential can have any amplitude, including a macroscopic one, while in the standard quantum case it depends on the microscopic Planck's constant $\hbar$. Therefore such a system is expected to exhibit some quantum-like macroscopic properties of a new kind (though certainly not every aspects of a genuine quantum system). 

In the present paper we apply this concept to surface gravity waves in an incompressible fluid. In this case we can build neither a wave function nor a quantum potential from the density of the fluid because of its constancy. However, one can write in this case a system of hydrodynamic equations in terms of a continuity equation and of a two-dimensional Euler equation where the density is replaced by the height profile of the fluid surface (see, e.g., \cite{Landau6}). Our new proposal is therefore to build a quantum-like potential, and then a wave function, from the square root of the surface height. One can then show that the gravity waves at the surface of a fluid subjected to such a generalized `quantum potential' would be solutions of a nonlinear Schr\"odinger equation whose form is typical of superfluids. Then we suggest to achieve this quantum potential in a real experiment, either by the action of a force at the surface (by wind, blower, electromagnetic field for a charged liquid, etc...), or by a continuous and adaptive deformation of the ground of the basin that contains the fluid.

In Section \ref{sec2} of this paper, we recall how one may construct a Schr\"odinger equation from the hydrodynamics equations of an irrotational fluid subjected to a generalized quantum-like potential. In Section \ref{sec3} , we write the equations of a gravity wave at the surface of a liquid subjected to such a quantum potential.  In Section \ref{sec4} , we apply these equations to the particular case of a quantized vortex in a basin with flat ground, which we compare to the classical vortex. In this case the quantum-like force is simulated by a classical force applied on the surface. In Section \ref{sec5} , we show that a curved ground amounts to a potential in the surface profile equation, so that a quantum-like potential can be directly applied by a specific deformation of the basin ground. This is once again exemplified by the case of the quantized vortex. We conclude in Section \ref{sec6}  by a short description of the experiments proposed to validate these concepts. 

%2*****************
\section{Theory}
%*****************
\label{sec2}

%*********************************************
\subsection{Pressureless approximation}
%********************************************

We consider a classical macroscopic compressible fluid described by the Euler and the continuity equations, in which we first assume that the pressure term is negligible,
\begin{equation}
\label{AA1}
 \l(\frac{\partial}{\partial t} + V \cdot \nabla\r) V  = -\nabla \phi,
\end{equation}
\begin{equation}
\label{AA2}
\frac{\partial \rho}{\partial t} + {\rm div}(\rho V) = 0,
\end{equation}
where $\phi$ is an exterior scalar potential. We assume that the fluid motion is potential, namely,
\beq
V=\nabla \varphi.
\eeq
We now assume that we apply to the fluid (using density measurements and a retroaction loop) a varying force which is a function of the fluid density in real time, namely, a  `quantum-like' force $F_Q$ deriving from the generalized quantum potential
\beq
Q=-2{\cal D}^2 \frac{\Delta \sqrt{\rho}}{\sqrt{\rho}}.
\eeq
It is a generalisation of the standard quantum potential \cite{Madelung,Bohm}, for which the constant $\cal D$ is restricted to the value ${\cal D}=\hbar/2m$, while here it can have any value. As recalled in what follows, this generalization still allows to recover a Schr\"odinger-type equation for a wave function $\psi$ subjected to Born's principle (i.e., the density is given by $\rho \propto |\psi|^2$) \cite{LN93}.

The Euler and continuity equation system becomes
\begin{equation}
\label{BBB1}
 \l(\frac{\partial}{\partial t} + V \cdot \nabla\r) V  = -\nabla \l({\phi}-2{\cal D}^2 \frac{\Delta \sqrt{\rho}}{\sqrt{\rho}}\r),
\end{equation}
\begin{equation}
\label{BBB2}
\frac{\partial \rho}{\partial t} + {\rm div}(\rho V) = 0,
\end{equation}
The system of equations (\ref{BBB1},\ref{BBB2}) can then be integrated under the form of a generalized Schr\"odinger equation. 

Indeed, equation~(\ref{BBB1}) takes the successive forms
\beq
\frac{\d}{\d t}(\nabla \varphi) +\frac{1}{2} \nabla (\nabla \varphi)^2+ \nabla \l({\phi}-2{\cal D}^2 \frac{\Delta \sqrt{\rho}}{\sqrt{\rho}}\r)=0,
\eeq
\beq
\nabla \l( \frac{\d \varphi}{\d t} +\frac{1}{2}(\nabla \varphi)^2+ {\phi}-2{\cal D}^2 \frac{\Delta \sqrt{\rho}}{\sqrt{\rho}}\r)=0,
\eeq
which can be integrated as
\beq
 \frac{\d \varphi}{\d t} +\frac{1}{2}(\nabla \varphi)^2+ {\phi}+K-2{\cal D}^2 \frac{\Delta \sqrt{\rho}}{\sqrt{\rho}}=0,
 \eeq
where $K$ is a constant that can be renormalized by a redefinition of the potential energy $\phi$. Let us now combine this equation with the continuity equation as follows:
\beq
\label{CCC}
\l[-\frac{1}{2} \sqrt{\rho}\l( \frac{\d \varphi}{\d t} +\frac{1}{2}(\nabla \varphi)^2+ {\phi}-2{\cal D}^2 \frac{\Delta \sqrt{\rho}}{\sqrt{\rho}}\r) + i \frac{\cal D}{2\sqrt{\rho}}\l( \frac{\partial \rho}{\partial t} + {\rm div}(\rho \nabla \varphi)  \r) \r] e^{i \varphi/2{\cal D}}=0.
\eeq
Finally we set
\beq
\label{psi}
\psi=\sqrt{{\rho}} \times e^{{i \varphi}/{2 {\cal D}}},
\eeq
and  the equation~(\ref{CCC}) is strictly identical to the following generalized Schr\"odinger equation:
\beq
{\cal D}^2 \Delta \psi + i {\cal D} \frac{\partial}{\partial t} \psi - \frac{\phi}{2}\psi = 0,
\eeq
as can be checked by replacing in it $\psi$ by its expression (\ref{psi}). Given the linearity of the equation obtained, one can normalize the modulus of  $\psi$ by replacing the matter density $\rho$ by a probability density  $P=\rho/M$, where $M$ is the total mass of the fluid in the volume considered: this will be equivalent.

The imaginary part of this equation amounts to the continuity equation, while its real part is the energy equation,
\beq
E=- \frac{\d \varphi}{\d t} =\frac{1}{2}V^2+ {\phi}-2{\cal D}^2 \frac{\Delta \sqrt{\rho}}{\sqrt{\rho}}.
\eeq

%*********************************
\subsection{Account of pressure}
%*********************************
Consider now the Euler equations with a pressure term and a quantum potential term: 
\begin{equation}
\label{AAA1}
 \l(\frac{\partial}{\partial t} + V \cdot \nabla\r) V  = -\nabla \phi-\frac{\nabla p}{\rho}+2{\cal D}^2 \,\nabla \l( \frac{\Delta \sqrt{\rho}}{\sqrt{\rho}}\r).
\end{equation}
In the case when ${\nabla p}/{\rho}=\nabla w$ is itself a gradient, its combination with the continuity equation can be still integrated in terms of a Schr\"odinger-type equation \cite{LN97},
\beq
{\cal D}^2 \Delta \psi + i {\cal D} \frac{\partial}{\partial t} \psi - \frac{\phi+w}{2}\psi = 0,
\eeq
Now the pressure term needs to be specified through a state equation, which can be chosen as taking the general form $p=k_p \rho^{\gamma}$.

In particular, in the sound approximation, the link between pressure and density writes $p-p_0=c_s^2(\rho-\rho_0)$, where $c_s$ is the sound speed in the fluid, so that $\nabla p/\rho=c_s^2 \nabla \ln \rho$. In this case, which corresponds to $\gamma=1$, we obtain the non-linear Schr\"odinger equation
\beq
{\cal D}^2 \Delta \psi + i {\cal D} \frac{\partial}{\partial t} \psi -k_p \ln |\psi| \; \psi =\frac{1}{2} \, \phi  \; \psi,
\eeq
with $k_p=c_s^2$. When $\rho-\rho_0 << \rho_0$, one may use the additional approximation $c_s^2 \, \nabla \ln \rho \approx (c_s^2 /\rho_0) \nabla \rho$, and the non-linear Schr\"odinger equation takes a form typical  of superfluids,
\beq
{\cal D}^2 \Delta \psi + i {\cal D} \frac{\partial}{\partial t} \psi - \beta |\psi|^2 \psi= \frac{1}{2} \, \phi  \; \psi,
\eeq
with $\beta={c_s^2}/{2 \rho_0}$. In the highly compressible case the dominant pressure term  is rather  $p \propto \rho^2$, and one obtains a non-linear Schr\"odinger equation of a similar form (see e.g. \cite{Nore,NoreThese}).

%3**************************************************************
\section{Gravity waves at the surface of a fluid (flat ground)}
\label{sec3}
%**************************************************************

%******************************************
\subsection{Schr\"odinger-type equation}
%******************************************

At first sight, gravity surface waves could be considered as unable to come under such a description, since they correspond to the case of an incompressible fluid. The constancy of the density $\rho$ prevents in this case to define a wave function from it (since $\rho$ is expected to be the square of its modulus) and to define a quantum potential (since it is given by second derivatives of $\rho$).   

However, we shall show that completely similar equations can be obtained for gravity waves by replacing the density by the height profile of the fluid surface. This allows us to suggest here a laboratory experiment in which a quantum-like force would be applied to the surface of the fluid and computed from the shape itself of this surface, thus forcing a superfluid-type behavior of the gravity waves.  In some respects, e.g., as concerns the wave behavior, this case is actually very similar to the case of sound (that we shall consider in a separate work), except for its two-dimensional character. But it has also the advantage to allow the possibility of fluctuations of the surface which are not small with respect to the height of the basin, and which may even reach $h=0$, therefore simulating with an uncompressible fluid the case of a perfectly compressible fluid.

We assume, as a first step, that the bottom of the basin is flat, that the average height of the fluid is $h_0=$cst, and that the effective height at a point of coordinate $(x,y)$ is 
\beq
h(x,y,t)=h_0+ \zeta(x,y,t).
\eeq
We consider an incompressible fluid, so that the variation of pressure involves a variation of its level in the basin. This implies a continuity equation in which the density is replaced by the fluid height $h$ \cite{Landau6}, namely,
\beq
\frac{\d h}{\d t} + \frac{\d}{\d x}( h \, v_x)+\frac{\d}{\d y}( h \, v_y)=0.
\eeq
We also assume that the velocities $v_x$ and $v_y$ do not depend on the $z$ coordinate and that they derive from a potential $\varphi$
\beq
v_x=v_x(x,y,t)=\frac{\d \varphi}{\d x}, \;\;\; v_y=v_y(x,y,t)=\frac{\d \varphi}{\d y}.
\eeq
Therefore the Euler equations for the two-dimensional velocity read
\beq
\frac{\d v_x}{\d t} +v_x \frac{\d v_x}{\d x}+v_y\frac{\d v_x}{\d y}=-\frac{1}{\rho} \frac{\d p}{\d x}-\frac{\d Q}{\d x},
\eeq
\beq
\frac{\d v_y}{\d t} +v_x \frac{\d v_y}{\d x}+v_y\frac{\d v_y}{\d y}=-\frac{1}{\rho} \frac{\d p}{\d y}-\frac{\d Q}{\d y},
\eeq
where $Q$ is the quantum potential to be applied to the fluid. Its expression will be specified in the following.
These equations can be completed by the energy equation (integral of Euler equations):
\beq
\frac{\d \varphi}{\d t}+ \frac{1}{2}(v_x^2+v_y^2+v_z^2)+Q+g z=- \frac{p}{\rho},
\eeq
where $g$ is the gravity acceleration.
At the surface of the fluid the pressure is constant (given by the atmospheric pressure for an open basin), so that the  right-hand side of this equation is a constant $- p_0/{\rho}$ which can be set to zero by a redefinition of the potential. Therefore the energy equation reads on the surface $z=h(x,y,t)$
\beq
\frac{\d \varphi}{\d t}+ \frac{1}{2}(v_x^2+v_y^2+v_z^2)+Q+g h=0.
\eeq
Assuming that $v_z=v_z(x,y,t)$, we can now separate the $(x,y)$ behavior and the vertical behavior ($z$) by combining the gravity term and the vertical velocity term under the form of a potential energy
\beq
\Phi(x,y,t)=g h+ \frac{1}{2} v_z^2,
\eeq
so that the two-dimensional continuity and energy equations now read
\beq
\frac{\d h}{\d t} + \frac{\d}{\d x}( h \, v_x)+\frac{\d}{\d y}( h \, v_y)=0,
\eeq
\beq
\frac{\d \varphi}{\d t}+ \frac{1}{2}(v_x^2+v_y^2)+Q+\Phi=0.
\eeq
Therefore, by taking the derivative of the energy equation we find that the Euler equations for the two-dimensional velocity finally take the form
\beq
\frac{\d v_x}{\d t} +v_x \frac{\d v_x}{\d x}+v_y\frac{\d v_x}{\d y}=-\frac{\d \Phi}{\d x}-\frac{\d Q}{\d x},
\eeq
\beq
\frac{\d v_y}{\d t} +v_x \frac{\d v_y}{\d x}+v_y\frac{\d v_y}{\d y}=-\frac{\d \Phi}{\d y}-\frac{\d Q}{\d y}.
\eeq
We recognize here the same form of the continuity and Euler equations that allowed us to combine them into a Schr\"odinger equation, but now with the density of matter $\rho$ replaced by the basin height $h$. This leads us to introduce a quantum potential that is a function of $\sqrt{h}$ instead of $\sqrt{\rho}$, given by 
\beq
Q=-2 {\cal D}^2 \; \frac{\Delta_2\sqrt{h}}{\sqrt{h}},
\eeq
where $\Delta_2={\d^2}/{\d x^2}+{\d^2}/{\d y^2}$. We are brought back to exactly the same situation as in the previous Sec.~\ref{sec2}, where the real continuity equation and energy equation may be combined in terms of a unique complex Schr\"odinger-type equation. Therefore we define a wave function $\psi(x,y,t)$ whose modulus is now the square root of the surface height profile while its phase is the two-dimensional velocity potential:
\beq
\psi = \sqrt{h} \times e^{i \varphi/2{\cal D}}.
\eeq
This wave function is solution of a generalized nonlinear Schr\"odinger equation,
\beq
{\cal D}^2 \Delta_2 \psi + i {\cal D} \frac{\partial  \psi }{\partial t}=\frac{1}{2} \, \Phi \,\psi =\frac{1}{2} \l(g \, h+ \frac{1}{2} \,v_z^2 \r) \psi.
\eeq
The vertical velocity term is non negligible only when it is larger than a critical value given by
\beq
g\, h= \frac{1}{2} \l(  \frac{d h}{dt} \r)^2,
\eeq
which is but the free fall equation, of solution
\beq
h_{\rm crit}=\frac{1}{2} \,g \, t^2, \;\;\; (v_z)_{\rm crit}=\pm g \,t.
\eeq
It can therefore be neglected in most real experimental situations, so that the equation  obtained, namely,
\beq
{\cal D}^2 \Delta \psi + i {\cal D} \frac{\partial  \psi}{\partial t}-\frac{1}{2} \,g  |\psi|^2 \,\psi=0,
\label{schro}
\eeq
is finally a two-dimensional macroscopic analog of the standard nonlinear Schr\"odinger equation of a superfluid (see e.g. \cite{Nore,Fetter1965}). But this quantum fluid-type behavior can now be macroscopic from the very beginning, since the parameter $\cal D$ is no longer constrained to its standard quantum mechanical value ${\cal D} = \hbar/2m$ as in the real superfluids known up to now.
 
 This does not mean that all quantum properties of a genuine superfluid will manifest themselves in such an experiment. In particular, one does not expect to observe the property of superfluidity itself (i.e., vanishing of viscosity), since the real viscosity of the classical fluid is unaffected. But some quantum-like properties like quantization, increase of the zone of potential motion, and possibly decrease of an effective viscosity, can nevertheless be expected.

%*****************************
\subsection{Wave equation in the linear case}
%***************************
Let us briefly consider the linearized case when the amplitude of the wave remains small compared with the height of the basin, i.e., $\zeta \ll h_0$. In this case the quantum potential can be simplified to lowest order as
\beq
Q= -\frac{{\cal D}^2 }{h_0} \Delta \zeta(x,y,t),
\label{QP}
\eeq
(which amounts to a two-dimensional Laplacian $\Delta_2$), while the wave function becomes
\beq
\psi=\sqrt{h_0} \l( 1+ \frac{1}{2} \frac{\zeta}{h_0} \r) \times e^{i \varphi/2{\cal D}}.
\eeq
In this case the equation of motion can be given the ``classical" form of a wave equation, but including a source term coming from the quantum potential. 

Before deriving this wave equation, let us first make a remark. Accounting for the fact that, at the surface of the fluid, 
\beq
\frac{\d h}{\d t}= v_z=\frac{\d \varphi}{\d z},
\eeq
and taking the derivative of the energy equation, we obtain (on the fluid surface):
\beq
\frac{\d^2 \varphi}{\d t^2}+ g \frac{\d \varphi}{ \d z}= -\frac{1}{2} \frac{\d }{\d t}(v^2)- \frac{\d Q}{\d t}.
\eeq
This shows that, contrarily to the sound case, there is no standard wave equation for the velocity potential.

However a generalized wave equation can be constructed for the surface profile. Indeed, the continuity equation takes  for $\zeta \ll h_0$ the simplified form
\beq
\frac{\d \zeta}{\d t} +h_0 \l( \frac{\d v_x}{\d x}+\frac{\d v_y}{\d y} \r)=0.
\eeq
Taking its time derivative and introducing the velocity potential, we obtain
\beq
\frac{\d^2 \zeta}{\d t^2} +h_0 \, \Delta_2  \l(\frac{\d \varphi}{\d t} \r)=0.
\eeq
Now the velocity potential is given by the energy equation
\beq
\frac{\d \varphi}{\d t}=-\l( \frac{1}{2}v^2+Q+g \, h\r),
\eeq
so that we finally obtain the wave equation of a classical wave  of propagation velocity $U= \sqrt{g h_0}$, but with an additional nonlinear term,
\beq
\frac{\d^2 \zeta}{\d t^2} -g  h_0 \, \Delta \zeta=\Delta\l(\frac{v^2}{2} + Q\r).
\eeq
By neglecting the $v^2$ term and by replacing the quantum potential by its linearized expression Eq.~(\ref{QP}), it finally becomes
\beq
\frac{\d^2 \zeta}{\d t^2} -g h_0 \, \Delta \zeta= -\frac{{\cal D}^2 }{h_0}\Delta^2 \zeta.
\eeq
This equation is structurally identical (after the reduction from 3 to 2 dimensions and the replacement of the density $\rho$ by the surface profile $\zeta$) to the wave equation directly obtained by linearization from the nonlinear Schr\"odinger equation of a superfluid \cite{NoreThese}.

%4%%%%%%%%%%%%%%%%%%%
\section{Application to a macroquantum vortex (flat ground and external force)}
\label{sec4}
%%%%%%%%%%%%%%%%%%%%

A particularly relevant example of application of  these ideas consists of the rotation of a fluid in a bucket with flat ground. We shall now show that it is possible to transform a classical vortex into a `macroquantum' superfluid-like vortex by application of a quantum-like force (e.g., by the action of a wind or of a blower) at the surface.

Let us first compare in detail the classical stationary vortex with the quantum vortex (see also \cite{NoreThese}).

\subsection{Classical vortex}\label{classic}
%***************************************cf LN27,99,139

\subsubsection{Height-velocity relation}
%********************************************

Let us use cylindrical coordinates $(r,\theta,z)$ (we denote the angle by $\theta$ instead of the usual notation $\varphi$ which have been used here for the velocity potential).  We consider the circular stationary flow of an uncompressible liquid with viscosity coefficient $\nu$ contained in a cylindrical bucket placed in the gravity field of the Earth, such that the velocity field has neither radial nor vertical  component, i.e., $v_r=0$ and $v_z=0$. The velocity field of such a vortex reduces to $v=v_{\theta}(r)$. In this case, the continuity equation is identically null and the three Navier-Stokes equations read \cite{Landau6}
\beq
\frac{v_{\theta}^2}{r}= \frac{1}{\rho} \,  \frac{ \d p }{\d r} ,
\eeq
\beq
\nu  \; \l(  \frac{\d^2 v_\theta}{\d r^2} + \frac{1}{r} \frac{\d v_\theta}{\d r}- \frac{v_\theta}{r^2 }  \r)=0,
\label{viscous}
\eeq
\beq
 \frac{1}{\rho}\,  \frac{ \d p }{\d z} + g=0.
\eeq
The first and third equations are solved in terms of a pressure expression that reads
\beq
p= p_0 +\rho \int \frac{v_{\theta}^2}{r} \, dr - \rho g z.
\eeq
Since the pressure is constant at the surface $z=h(r)$, the surface profile is given by
\beq
h=h_0+ \frac{1}{g} \, \int \frac{v_{\theta}^2}{r} \, dr.
\label{equ49}
\eeq
Reversely, by derivating this expression, we obtain a general relation between the velocity and the height profile,
\beq
\label{equ59}
v_{\theta}^2 (r)=g\, r\, \frac{ \d h}{ \d r}.
\eeq

\subsubsection{Height and velocity profiles}
%**********************************************

The second equation (\ref{viscous}) is solved as
\beq
v_\theta= \alpha \, r + \frac{\beta}{r}.
\eeq
Therefore, one finds from Eq.~(\ref{equ49}) that the surface height reads
\beq
h=h_0+ \frac{1}{g} \l[ \frac{1}{2} \,\alpha^2 \, r^2 + 2 \alpha \beta \ln \frac{r}{r_1}- \frac{1}{2} \frac{\beta^2}{r^2} \r].
\eeq
The velocity solutions are the same as that found for Couette flows between two rotating cylinders.  But more general solutions are obtained for flows described by different values of the integration constants in different zones that are matched by writing the continuity of $h$ and of $v_\theta$, and therefore also of the derivative of $h$.

One recognizes, in the two components of the velocity in this solution, the potential rotation $v_\theta={\beta}/{r}$ with hyperbolic surface profile, $h=h_0(1-r_0^2/r^2)$, and the solid rotation case $v_\theta= \alpha \, r$ with the parabolic surface profile of Newton's rotating bucket, $h \propto r^2$. In particular, since the divergent terms should vanish when $r\to 0$, one finds that $\beta=0$, so that the height profile must be parabolic in the central region which rotates as a solid body due to viscosity.

Now, the strictly quantum vortex is known to be everywhere potential. We shall therefore compare it with a particular case of the classical vortex which is potential in its outer regions. 

 Let us call $r_m$ the limit between the inner region where solid-like rotation $v_\theta=\Omega r$ occurs and the outer region with rotation velocity  $\propto 1/r$, and $R$ the bucket radius. Let us also assume that the bucket is itself in rotation with a velocity $v_R= \Omega \,r_m^2/R$. This choice has the advantage to preserve the exact $1/r$ rotation in the outer region, as in an infinite radius vortex. 
 
Let $\Gamma= 2 \pi \Omega r_m^2$ be the circulation around the vortex and $h_0$ be now defined as the asymptotic height for $r \to \infty$. The fluid height at center is given by
 \beq
 h_c=h_0 - \frac{ \Gamma \, \Omega}{ 2 \pi g}.
 \eeq
Still with the aim to compare a classical vortex with the quantum vortex, we consider in what follows only  the case when the height is zero at center, $h_c=0$. In the inner region, the velocity and surface height profile are
\beq
 v_\theta({\rm in})= \Omega r, \;\;\; h({\rm in})= \frac{\Omega^2}{2g} \,r^2.
\eeq
In the external region the motion is potential and such that
\beq
 v_\theta({\rm ex})= \frac{\Omega\, r_m^2}{r}, \;\;\; h({\rm ex})=h_0 \l \{1-\l(\frac{r_0}{r}\r)^2 \r\}.
\eeq
We verify that $v_\theta= \Omega r_m$ on the matching radius $r_m$ from both inner and outer relations. The radius $r_0$ which appears in the external profile corresponds to the extrapolation of this profile to zero height, $h=0$. The matching condition yield 
\beq
r_m^2=2\, r_0^2.
\eeq
The asymptotic height $h_0$ is given from the liquid height $h_R$ at bucket radius $R$ by the relation
\beq
h_0= \frac{h_R}{1-(r_0/R)^2}.
\eeq
The circulation $\Gamma= 2 \pi \Omega r_m^2$ is given in this case ($h=0$ at center) by
\beq
\Gamma=4 \pi \Omega r_0^2 =\frac{2 \pi g h_0}{\Omega},
\eeq
so that one obtains the relation (which can be directly obtained from Eq.~(\ref{equ59}))
\beq
r_0^2= \frac{g h_0} {2 \,\Omega^2}. 
\eeq 
The outer velocity therefore reads, in terms of the experimentally chosen values of $h_0$ and $\Omega$,
\beq
 v_\theta({\rm ex})= \frac{g h_0}{\Omega} \; \frac{1}{r}.
\eeq
By matching the inner and outer profiles at radius $r_m$, one finds that $h_m=h_0/2=\Omega^2 r_0^2/g$, which yields again the same relation. The corresponding height profile is given in Fig.~\ref{profile}.

\subsection{Quantum-like vortex}\label{qlv}
%***************************************cf LN27,130

\subsubsection{Generalized height-velocity relation}
%********************************************
We have considered, in Sec.~\ref{classic}, the standard classical vortex in a flat ground bucket. Let us now assume that we apply to the same fluid a quantum-like potential $Q(r)$, which is a function of $r$ only and depends on the surface profile. It is applied by the action of a classical force on the surface itself (e.g., by a wind, a blower, or any other mean, pressure field, electromagnetic field, etc...) that simulates a quantum-like force. Let us set
\beq
f=\sqrt{\frac{h}{h_0}},
\eeq
where $h_0$ is a reference height for the fluid in the basin (its precise definition will emerge in the following). The quantum potential reads
\beq
Q=-2 {\cal D}^2 \;  \frac{\Delta f}{f}= \frac{-2 {\cal D}^2}{f}  \l( \frac{\d^2 f}{\d r^2} + \frac{1}{r} \frac{\d f}{ \d r} \r) .
\eeq
We still consider the purely circular velocity case. Let us first assume that the viscosity is negligible. The Euler equations with quantum potential become
\beq
\frac{v_{\theta}^2}{r}= \frac{1}{\rho} \,  \frac{ \d p }{\d r}+ \frac{\d Q}{\d r},
\eeq
\beq
 \frac{1}{\rho}\,  \frac{ \d p }{\d z} + g=0,
\eeq
i.e.,
\beq
\frac{v_{\theta}^2}{r}= \frac{1}{\rho} \,  \frac{ \d p }{\d r} -2 {\cal D}^2 \frac{\d}{\d r} \l\{ \frac{1}{f} \l( \frac{\d^2 f}{\d r^2} + \frac{1}{r} \frac{\d f}{ \d r} \r)  \r\},
\eeq
\beq
 \frac{ \d p }{\d z} =-g \rho.
\eeq
This last equation can be integrated as $ p/\rho=p_0/\rho+g (h-z)$, since the pressure is constant on the free surface $z=h(r)$, so that one obtains the relation
\beq
 \frac{1}{\rho} \,  \frac{ \d p }{\d r} = g \, \frac{\d h}{\d r}.
\eeq
Knowing that $h=h_0\,f^2$, one can easily generalize to the quantum potential case the previous classical height-velocity relation (Eq.~\ref{equ59}):
\beq
v_{\theta}^2(r)= g \,r \frac{\d}{\d r}  \l\{ h- \frac{2 {\cal D}^2}{g\sqrt{h}} \l( \frac{\d^2\sqrt{h}}{\d r^2} + \frac{1}{r} \frac{\d \sqrt{h}}{ \d r}   \r)  \r\} .
\eeq

\subsubsection{Superfluid-like height and velocity field}
%********************************************************
Assume now that the fluid remains potential, at least in a large external part of the vortex. One may now construct, in this domain, a wave function from the surface height $h=h_0 f^2$ and from the potential of velocity $\varphi$, which reads  $ \psi = f(r) \, e^{i \varphi/2{\cal D}}$ and is solution of the nonlinear Schr\"odinger equation~(\ref{schro}). 

The single vortex solution of this equation is such that $\varphi= 2{\cal D} l \theta$ (see e.g. \cite{Fetter1965,Kawatra66,NoreThese}), so that the velocity field may be written as
\beq
v_{\theta}= \frac{1}{r}\, \frac{\d \varphi}{\d \theta}=\frac{2{\cal D} l}{r}.
\eeq
One recovers, even in this generalized quantum-like case, the $1/r$ behavior of a potential rotation velocity. Moreover, the viscosity terms in the Navier-Stokes equation (Eq.~\ref{viscous}) identically vanishes for such a flow, so that the NL schr\"odinger equation is also valid for a real viscous fluid, at least in the exterior part of the vortex (the inner heart being still expected to rotate like a solid body).

Since $v_{\theta}^2/r= -\d(2 {\cal D}^2 l^2/r^2)/\d r$, one obtains after integration,  for such a vortex, the equation
\beq
 \frac{\d^2 f}{\d r^2} + \frac{1}{r} \frac{\d f}{ \d r} + \l( \frac{1}{a^2}-\frac{ l^2}{r^2} \r) f - \frac{g h_0}{2{\cal D}^2}\, f^3=0,
\label{superflu}
\eeq
which has solutions only for integer values of $l$, i.e., for quantized values of the velocity. The constant $a$ is an integration constant whose value can be determined from the asymptotic infinite limit of the vortex. Indeed, when $r \to \infty$, we have by definition $h \to h_0$ so that $f \to 1$, while $\d f/ \d r \to 0$ and $\d ^2 f/\d r^2 \to 0$, so that we obtain
\beq
a^2=\frac{2{\cal D}^2}{g \, h_0}.
\label{cond}
\eeq
Let us compare this equation with that of a vortex in a real quantum superfluid \cite{Onsager,Feynman,Gross,Pitaevskii}. A liquid Helium II superfluid can be described, under certain conditions, by a non-linear Schr\"odinger equation
\beq
-\frac{\hbar^2}{2m}\, \nabla^2 \psi + V_0 |\psi|^2 \psi = \mu \psi,
\eeq
where $V_0$ characterizes a short range repulsive potential and $\mu$ is the chemical potential. The presence of a single quantized vortex in the fluid is characterized by a wave function of the type
\beq
\psi= f_l(r) \, e^{il \theta},
\eeq
where ($r,\theta$) are the appropriate cylindrical coordinates and $l$ an integer. The function $f_l(r)$ is then a solution of the equation \cite{Kawatra66}
\beq
 \frac{\d^2 f}{\d r^2} + \frac{1}{r} \frac{\d f}{ \d r} + \l( \frac{1}{a_s^2}-\frac{ l^2}{r^2} \r) f - \frac{1}{a_s^2}\, f^3=0.
\eeq
In this equation the fundamental distance scale which governs the vortex scale is but the de Broglie length $a_s$, which is given by \cite{Kawatra66}
\beq
a_s^2= \frac{\hbar^2}{2m\mu}.
\eeq
Therefore the macroquantum vortex equation (\ref{superflu}) has exactly the same form as this real quantum superfluid vortex equation. The distance scale $a$ given by the relation $a^2={2{\cal D}^2}/{g \, h_0}$, which characterizes the scale of the vortex, can be therefore identified, as expected, as a generalized de Broglie length. Indeed, we recover standard quantum mechanics in the special case $2{\cal D}=\hbar/2m$, so that, for $m=1$ which is relevant here, the gravity potential $g h_0$ in the expression $a=2{\cal D}/\sqrt{2 g \, h_0}$ plays a role similar to that of the chemical potential $\mu$ in the real superfluid case. 

Here, the square of the parameter ${\cal D}$ characterizes the amplitude of the macroquantum potential $Q$. We shall therefore be able to simulate some of the properties of a genuine quantum vortex (usually constrained by the microscopic value of $\hbar$) in a now macroscopic experiment.

%%%%%%%%cf Tourbillon.nb
\begin{figure}[!ht]
\begin{center}
\includegraphics[width=12cm]{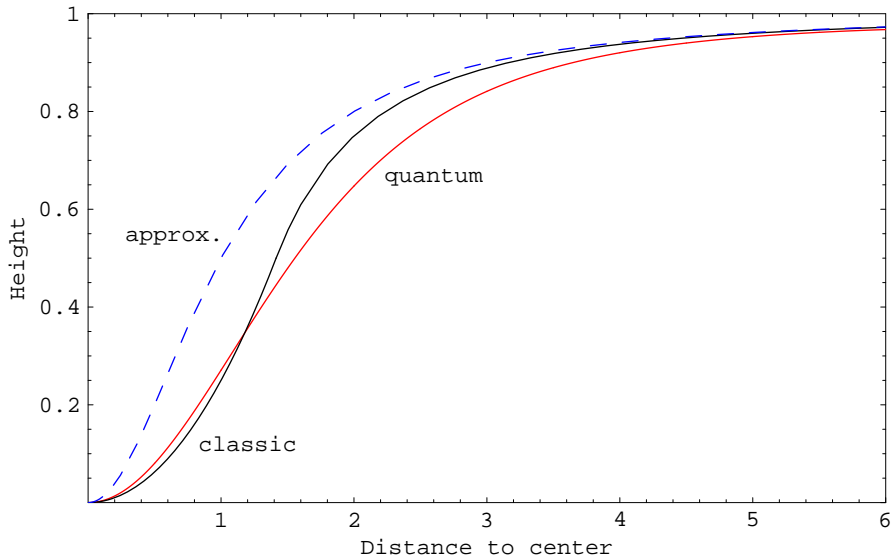}
\caption{\small Comparison, for $l=1$, between the height profiles of a quantum-like vortex (`quantum'), its classical counterpart (`classic') for the same limiting conditions at edge $r=6a$, and Fetter's approximation to the quantum vortex (dashed line). The radial distance to center $r$ is in unit of $a=r_0$ (see text).}
\label{profile}
\end{center}
\end{figure}
%%%%%%%%

%%%%%%%%cf Tourbillon.nb
\begin{figure}[!ht]
\begin{center}
\includegraphics[width=12cm]{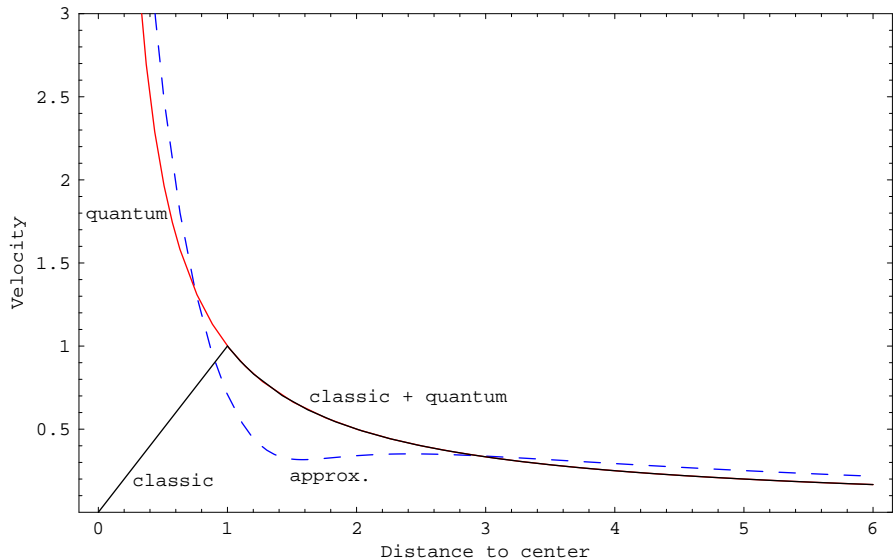}
\caption{\small Comparison between the velocities of a quantum superfluid vortex with $l=1$ (`quantum'), its classical counterpart (`classic') for the same limit conditions at $r=6 a$, and Fetter's approximation to the quantum vortex (dashed line).}
\label{vitesse}
\end{center}
\end{figure}
%%%%%%%%

The quantum vortex profile is given by the solution of Eq.~(\ref{superflu}). An analytical approximation to this solution has been given by Fetter \cite{Fetter1965}, 
as $h_F(r)=h_0 \; {r^2}/{(r_0^2+r^2)}$,
but it is not precise enough for our purpose (see Fig.~\ref{profile}). In the central region, the solution is a Bessel function $f(r) \propto J_l(r/r_0)\sim r^l$ \cite{Kawatra66} (it is remarkable that the central profile is therefore parabolic when $l=1$).
We give in Fig.~\ref{profile} the result of a numerical integration of Eq.~(\ref{superflu}) in the case $l=1$, which agrees with those of Kawatra \cite{Kawatra66} and Nore \cite{NoreThese}.

The comparison between the quantum-like and the classical vortex profiles (Fig.~\ref{profile}) shows significant differences. The quantum height profile is higher than the classical one in the central region and lower in the outer region, and they differ in both regions by a relative value that reaches $\approx 20 \%$, in a manner that should be observable.

The quantum vortex velocity is given by $v_\theta(r)=2{\cal D} l/r$ everywhere, and it is therefore quantized.  In the Fetter approximation, one can obtain it from the above velocity-height relation, but the result is once again not precise enough (see Fig.~\ref{vitesse}). The continuation of an $1/r$ velocity, i.e. of a potential motion in the inner region of the vortex and its quantization achieve two additional signatures of a quantum superfluid-like behavior which can be put to the test in a real experiment, even though, due to viscosity, one does not expect it to be achieved for all values of the radius, contrarily to the genuine quantum superfluid.

\subsubsection{Quantum force applied to the surface}
%***************
Let us finally give the expression of the quantum potential, and therefore that of the quantum force that derives from it, to be applied to the fluid surface in the proposed experiment. They can be directly established from Eq.~(\ref{superflu}) as a function of the surface height profile, namely, since $2{\cal D}^2/a^2=g h_0$,
\beq
Q= \frac{-2 {\cal D}^2}{f}  \l( \frac{\d^2 f}{\d r^2} + \frac{1}{r} \frac{\d f}{ \d r} \r)=2 {\cal D}^2 \l( \frac{1}{a^2}-\frac{ l^2}{r^2} \r) -g \, h=g h_0 \l( 1- l^2 \frac{a^2}{r^2}-\frac{h}{h_0} \r),
\eeq
and
\beq
F_Q=-\frac{\d Q}{\d r}= \frac{-4 {\cal D}^2 l^2}{r^3} + g \, \frac{\d h}{\d r}= g h_0 \l(- \frac{2a^2 l^2}{r^3}+ \frac{1}{h_0}\frac{\d h}{\d r} \r).
\eeq
We note that the existence of the quantum potential cancels the gravity term $g \, h$ in the total potential energy, which allows the velocity to remain everywhere proportional to $1/r$ in the quantum-type case.

%%%%%%%%cf Tourbillon.nb
% QuantPot.eps était faux d'un facteur 2 A refaire
%voir si on remet cette figure etautres valeurs de l
%%%%%%%%

%*5**************************************LN27,198-205
\section{Generalization to a curved ground}
\label{sec5}
%***************************************

\subsection{Simulation of a quantum potential by a non-flat ground}
%*******************************************
In the previous study, we have assumed that the ground of the basin containing the liquid was flat. Let us now consider a more general case, namely, that of a basin with a non-flat ground, which will reveal itself to be particularly relevant for a explicit application of a quantum potential to a liquid in a real experiment,

Let $h_b=h_b(r,\theta,t)$ denote the height profile of the basin ground, while $h(r,\theta,t)$ still denotes the height of the fluid surface (measured with respect to a constant level). In this case the continuity equation no longer holds in terms of the visible surface height $h$, but instead in terms of the fluid height,
\beq
h_q= h-h_b.
\eeq 
Namely, it reads
\beq
\frac{\d h_q}{\d t} + \frac{\d}{\d x}( h_q \, v_x)+\frac{\d}{\d y}( h_q \, v_y)=0.
\eeq
This means that it is now $\sqrt{h_q}$ instead of $\sqrt{h}$ that can be called to play the role of a wave function modulus. When $v_z=0$, the vertical Euler equation reads
\beq
\frac{1}{\rho} \frac{\d p } {\d z} +g=0,
\eeq
so that we recover the same expression for the pressure as in the flat ground case. Namely, let $p=p_0$ be the constant pressure at the free surface $z=h$ (e.g., the atmospheric pressure), the pressure therefore reads \cite{Landau6}
\beq
\frac{p}{\rho}= \frac{p_0}{\rho}+ g\,(h-z).
\eeq 
The Euler equation therefore keeps the form of the constant ground case. Namely, the pressure term in this equation remains $-\nabla p/\rho=-g \, \nabla h$, i.e., it still depends on the surface height $h$ while the continuity equation now depends on the fluid height $h_q$. 

But let us now decompose the surface height as $h=h_q+h_b$. The
radial Euler equation (in cylindrical coordinates) becomes, in the absence of an additional external potential,
\beq
\frac{\d v_r}{\d t}= - v_r \frac{\d v_r}{ \d r}+ \frac{v_{\theta}^2}{r}-g \frac{\d h_q}{\d r} -g \frac{\d h_b}{\d r}.
\eeq
We have therefore recovered the previous form for the couple of Euler and continuity equations (now in cylindrical coordinates), but they are now written in terms of  $h_q$, while the $h_b$ term can be considered as an externally added potential. This result is remarkable since it means that the curved ground profile $h_b(r, \theta,t)$ now plays the role of a potential (up to the gravity constant $g$).

As we shall see in the following, this property can be used in at least two ways. (i) In natural systems:  search for natural sea grounds that would have the expected profile and could therefore locally create a quantum-like wave; (ii)  simulate the quantum potential by the profile of the basin ground. Namely, we propose a laboratory experiment in which one would simulate a quantum potential by an adequately curved ground. This can be done either by applying the theoretically expected potential for a stationary solution, or by simulating the time-dependent quantum behavior by continuously deforming, in an adaptive way, the ground height profile according to the relation
\beq
h_b=- \frac{2 {\cal D}^2}{g} \; \frac{\Delta_2 \sqrt{h_q}}{\sqrt{h_q}},
\eeq
in which the ground height becomes a function of the fluid height $h_q=h-h_b$. 

We can now, in the case of potential motion, build a wave function from the fluid height $h_q$ and the potential of the velocity field $\varphi$, 
\beq
\psi= \sqrt{\frac{h_q}{h_0}} \times e^{i \varphi/2{\cal D}},
\eeq
so that the gravity term $g h_q$ becomes a non linear term $g|\psi|^2$.  In this case the Euler and continuity equations can be combined in terms of the nonlinear Schr\"odinger equation which is typical of superfluids,
\beq
{\cal D}^2 \Delta_2 \psi + i {\cal D} \frac{\partial}{\partial t} \psi -\frac{1}{2} g \, |\psi|^2 \psi=0.
\eeq

\subsection{Application to a quantum-like vortex}
%*******************************************
Let us examplify this result, as in the flat ground case accompanied by a quantum-like force, by the study of a stationary vortex. 

In such an experiment, the `classical' case now corresponds to a flat ground $h_b=0$, and therefore to a situation where the surface height is $h=h_q+h_b=h_q$. The `macroquantum' case involves a basin ground $h_b\neq 0$ that plays the role of a quantum potential, and therefore a surface height $h=h_q+h_b \neq h_q$. One does not directly see $h_q= h_0|\psi|^2=h-h_b$, it is reconstructed from the measurement of $h$ and $h_b$. Figure~\ref{profile} therefore corresponds, in this case, to a comparison between the two shapes of $h_q= h-h_b$ in the ``classical" ($h_b=0$) and ``quantum-like" cases ($h_b=Q/g$).

The various results of Sec.~\ref{qlv} still apply in this case, after having replaced $h$ by $h_q$. In particular, the angular momentum remains quantized in terms of the quantum number $l=0,1,2,...$, the wave function reads
 \beq
 \psi=\sqrt{\frac{h_q}{h_0}} \times e^{i l \theta}
 \eeq
and the velocity field is therefore still quantized as
\beq 
v_\theta(r)= \frac{2{\cal D}l}{r}.
\eeq

\subsubsection{Curved ground height}
\begin{figure}[!ht]
\begin{center}
\includegraphics[width=12cm]{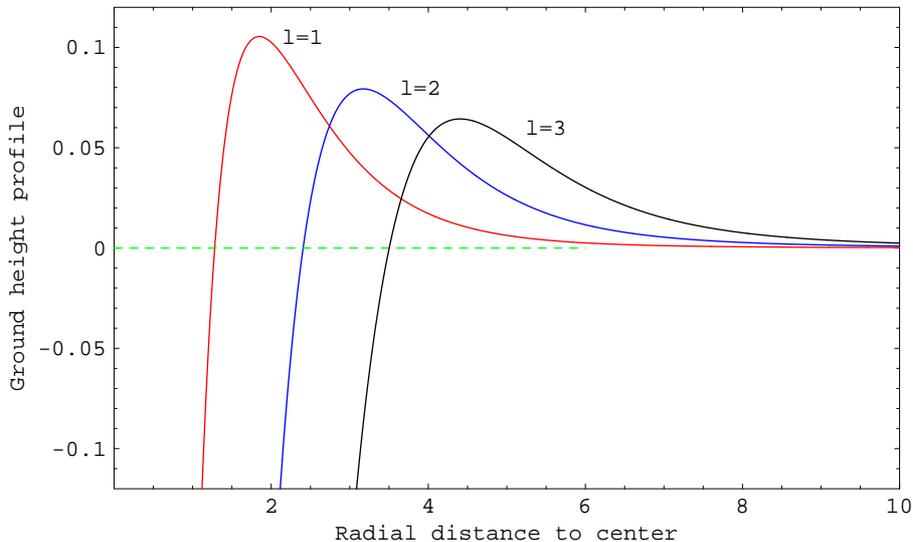}
\caption{\small Height profiles of the grounds that simulate a quantum-like potential, for three values of the quantum number $l$. The abscissa unit is the generalized de Broglie length $a= 2{\cal D} / \sqrt{2 g h_0}$, while the ordinate unit is the asymptotic surface height  $h_0$.}
\label{troisfonds}
\end{center}
\end{figure}

%************************************
Equation~(\ref{superflu}) now applies to $f_q=\sqrt{h_q/h_0}$ and reads
\beq
 \frac{\d^2 f_q}{\d r^2} + \frac{1}{r} \frac{\d f_q}{ \d r} + \l( \frac{1}{a^2}-\frac{ l^2}{r^2} \r) f_q -  \frac{1}{a^2}\, f_q^3=0,
 \label{superfond}
\eeq
where the characteristic generalized de Broglie scale $a$ is still given by
\beq
a^2=\frac{2{\cal D}^2}{g h_0}. 
\eeq
Let us now express the ground height 
\beq
h_b=-\frac{2{\cal D}^2}{g}\; \frac{\Delta_2 f_q}{f_q},
\eeq
which is the quantum potential up to the gravity constant $g$, in a dimensionless way. To this purpose, we replace the radial variable $r$ by the dimensionless variable $x=r/a$, and we obtain, thanks to the above expression for $a$, 
\beq
h_b=h_0 \, F_l(r/a),
\eeq
where the expressions in factor of $h_0$,
\beq
F_l(x)= -\frac{1}{f_q} \l(\frac{\d^2 f_q}{\d x^2} + \frac{1}{x} \frac{\d f_q}{ \d x}\r)
\eeq
are now purely numerical functions that depend only on the value of $l$. We have determined them by performing a numerical integration of the superfluid equation (which agrees with those of Refs.~\cite{Kawatra66,NoreThese}), and plotted them in Fig.~\ref{troisfonds}.

%%%%%%%%cf FondCourbe.nb
%%%%%%%%

\subsubsection{Free surface and liquid heights}
%*************
Let us now determine the free surface height profile in this `macroquantum' case. From the superfluid-like equation~(\ref{superfond}), one may write the ground height as
 \beq
h_b=-h_0\;\frac{1}{f_q} \l(\frac{\d^2 f_q}{\d x^2} + \frac{1}{x} \frac{\d f_q}{ \d x}\r)= -h_0\,f_q^2 +h_0\l( 1- l^2 \frac{a^2}{r^2}\r). 
\eeq
Finally, since $h_0f_q^2=h_q$ and $h=h_q + h_b$, one finds that the surface profile is given by
\beq
h_l(r)= h_0\l( 1- l^2 \frac{a^2}{r^2}\r).
\label{surf}
\eeq
It is noticeable that one obtains the correct asymptotic height $h_0$ for $r \to \infty$, which means that the relation (Eq.~\ref{cond}) could have been established directly from this condition. 

One therefore recovers the outer classical profile $h=h_0(1-r_0^2/r^2)$, with two fundamental differences:

\noindent (i) The values of the characteristic scale $r_0$ are now quantized in agreement with the quantization of the differential rotational velocity $\Omega=2{\cal D}l/r^2$, namely,
\beq
(r_{0})_l= \sqrt{\frac{2}{gh_0}}\; {\cal D} \,l.
\eeq
\noindent (ii) This profile is expected to continue in the inner region below the classical matching radius $r_m$. This agrees with the obtention of the same result for the velocity field (see Fig.~\ref{vitesse}), since this profile is typical of potential motion \cite{Guyon}. 

%%%%%%%%cf FondCourbe.nb
\begin{figure}[!ht]
\begin{center}
\includegraphics[width=12cm]{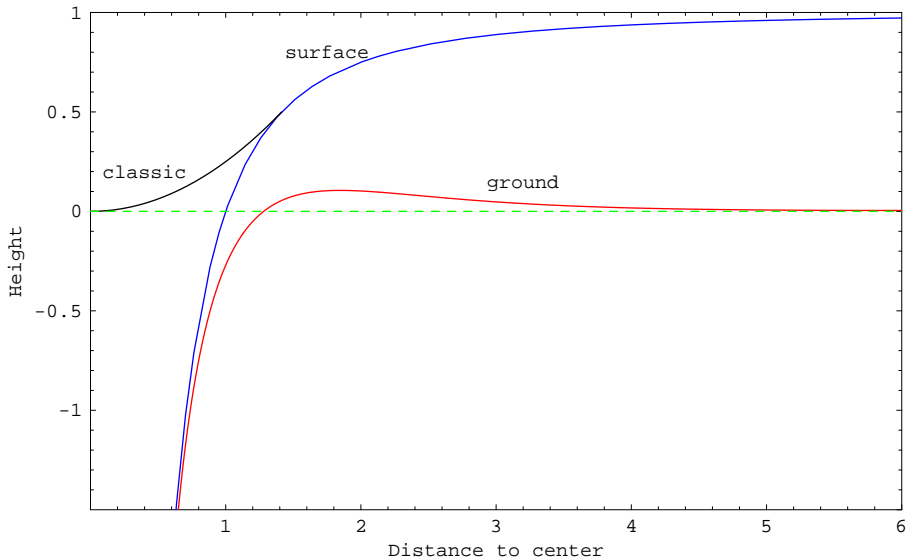}
\caption{\small Profiles of the curved ground $h_b$ that simulates a quantum potential and of the corresponding surface height $h$, compared with the `classical' surface profile (flat ground). The quantum-like profile, given in Fig.~\ref{profile} (for $l=1$), which is a solution of a nonlinear Schr\"odinger equation of the superfluid kind, is the liquid height given by the difference $h_q=h-h_b$.}
\label{ground}
\end{center}
\end{figure}
%%%%%%%%

%%%%%%%%
%*****************
\section{Discussion and conclusion}
\label{sec6}
%*****************

We have given in this paper the theoretical basis aiming at preparing a new kind of quantum-like laboratory experiment, in which the surface gravity waves of a classical liquid in a basin of finite height could be transformed to acquire some quantum-type properties typical of superfluids. Note that some preliminary numerical simulations have yielded a first validation of this concept in the three dimensional case \cite{NL06}, giving the hope that a real laboratory experiment should be possible to achieve.

In the classical hydrodynamic case considered in this work, possible shortcomings are to be considered, such as the effects of vorticity, of viscosity at small scales, of uncertainties in the measurement of the height and in the application of the quantum-like force or in the deformation of the basin ground, etc... 

The classical vortex is rotationnal in the inner region which rotates like a solid body, and possibly potential in its outer region (a zone that vanishes in the case of the rotationg bucket), while the quantum vortex (see, e.g., \cite{Fetter1965}) is everywhere potential. It is clear that, in the new macroscopic quantized vortex experiment, the potential motion is not expected to be preserved up to the center of the bucket, since the viscosity of the classical fluid implies the existence of a core radius inside which solid rotation is expected, and then vorticity. Therefore, although the fluid height is partly described by a superfluid equation, this does not mean that the specific property of superfluidity, namely, of the vanishing of viscosity, is expected in such an experiment, since the microscopic classical nature of the fluid is not affected. However, the region of potential motion is expected to be increased toward the center in a larger domain than for the standard classical fluid, so that we could simulate in this way the decrease of an effective viscosity.

We shall in forthcoming works attempt to take into account these effects in more complete theoretical analyses and numerical simulations with improved integration schemes  \cite{NL06}, then we intend to lead a real hydrodynamic laboratory experiment aiming at implementing the `macroquantum' behavior described in the present paper \cite{Nottale07B}. 

In addition to applications to experimental devices, the concept described in the present paper could also be applied to the understanding of `freak waves'. They are extraordinary large water waves whose heights exceed by a factor larger than 2 the usual wave height, and which may have potentially devastating effects. It has been established that the nonlinear Schr\"odinger equation can describe many of the features of their dynamics, although it still remains unclear how they are generated in realistic oceanic conditions (see \cite{Onorato2001} and references therein). The results obtained here suggest new possibilities of understanding some of these freak waves, which may naturally arise for certain combinations of the surface height profile and wind conditions, and/or of both the surface and ground profiles. These combinations, rare but not impossible due to the large number of configurations on the whole surface of the Earth oceans, would simulate the appearance of a quantum-like potential and would therefore lead to quantum-like waves, whose height can be doubled respectively to classical waves because it is given by the square of the modulus of a wave function. A future work will be specially devoted to this application.

Let us conclude by remarking that, provided the experiment proposed here succeeds, it could lead to many new applications in several domains: didactic ones (teaching of quantum mechanics), laboratory physics (macroscopic models of quantum systems, simulations of atomic and molecular systems, study of the quantum to classical transition, laboratory astrophysics \cite{LN96,NSG97,LN97,NSL00}, models of biological-like systems \cite{LN04,Auffray2007A,Nottale2007A}), new technology (development of new devices having some quantum-like properties and behavior), self-organization (plasma confinement, control of turbulence, etc..).\\

%%%%%%%%%%%%%%%%%%
{\bf Acknowledgments} I gratefully thank Drs. Thierry Lehner, M.N. C\'el\'erier and C. Auffray and for helpful discussions during the preparation of this paper.

%%%%%%%%%%%%%%%

%*****************
\end{document}